\documentclass[traditabstract]{aa}

\usepackage{natbib}
\bibpunct{(}{)}{;}{a}{}{,}

\usepackage{xspace}

\usepackage{graphicx}
\usepackage{float}

\newcommand{\microns}{\,$\mu$m\xspace}
\newcommand{\degrees}{$^\circ$\xspace}
\newcommand{\sqdeg}{\,deg$^2$\xspace}

\newcommand{\kpc}{\,kpc\xspace}

\begin{document}

\titlerunning{A self-consistent model of Galactic emission}
\authorrunning{Robitaille et al.}

\title{A self-consistent model of Galactic stellar and dust infrared emission and the abundance of polycyclic aromatic hydrocarbons}

\author{
  Thomas P. Robitaille\inst{\ref{inst:mpia}, \ref{inst:harvard}}
    \and
  Ed Churchwell\inst{\ref{inst:uw}}
    \and
  Robert A. Benjamin\inst{\ref{inst:uww}}
    \and
  Barbara A. Whitney\inst{\ref{inst:uw},\ref{inst:ssi}}
    \and
  Kenneth Wood\inst{\ref{inst:sta}}
    \and
  Brian L. Babler\inst{\ref{inst:uw}}
    \and
  Marilyn R. Meade\inst{\ref{inst:uw}}
}

\institute{
  Max Planck Institute for Astronomy, K\"onigstuhl 17, Heidelberg 69117, Germany
  \email{robitaille@mpia.de}
  \label{inst:mpia}
    \and
  Harvard-Smithsonian Center for Astrophysics, 60 Garden Street, Cambridge, MA, 02138, USA
  \label{inst:harvard}
    \and
  Department of Astronomy, University of Wisconsin-Madison, Madison, WI 53706
  \label{inst:uw}
    \and
  Department of Physics, University of Wisconsin-Whitewater, Whitewater, WI 53190
  \label{inst:uww}
    \and
  Space Science Institute, 4750 Walnut St. Suite 205, Boulder, CO 80301
  \label{inst:ssi}
    \and
  School of Physics and Astronomy, University of St. Andrews, North Haugh, St. Andrews, Fife, KY16 9SS, UK
  \label{inst:sta}
}

\abstract{
We present a self-consistent three-dimensional Monte-Carlo radiative transfer model of the stellar and dust emission in the Milky-Way, and have computed synthetic observations of the 3.6 to 100\microns emission in the Galactic mid-plane. In order to compare the model to observations, we use the GLIMPSE, MIPSGAL, and IRAS surveys to construct total emission spectra, as well as longitude and latitude profiles for the emission. The distribution of stars and dust is taken from the SKY model, and the dust emissivities includes an approximation of the emission from polycyclic aromatic hydrocarbons in addition to thermal emission. The model emission is in broad agreement with the observations, but a few modifications are needed to obtain a good fit. Firstly, by adjusting the model to include two major and two minor spiral arms rather than four equal spiral arms, the fit to the  longitude profiles for $|\ell|>30^\circ$ can be improved. Secondly, introducing a deficit in the dust distribution in the inner Galaxy results in a better fit to the shape of the IRAS longitude profiles at 60 and 100\microns. With these modifications, the model fits  the observed profiles well, although it systematically under-estimates the 5.8 and 8.0\microns fluxes. One way to resolve this discrepancy is to increase the abundance of PAH molecules by 50\% compared to the original model, although we note that changes to the dust distribution or radiation field may provide alternative solutions. Finally, we use the model to quantify which stellar populations contribute the most to the heating of different dust types, and which stellar populations and dust types contribute the most to the emission at different wavelengths.
}

\keywords{Radiative transfer -- Galaxy: structure --- Infrared: ISM}

\maketitle

\section{Introduction}

\label{sec:introduction}

Although it has been common to characterize the large-scale Galactic diffuse emission by treating each component of the Galaxy individually -- the neutral gas distribution, HII region maps, the magnetic field structure -- this emission is usually caused by a convolution of different combinations of these components. Far-infrared dust emission, for example, traces the dust distribution heated by stars; radio free-free emission depends specifically on the local density of main-sequence massive stars; radio synchrotron emission traces the local cosmic ray electron density (and energy spectrum) convolved with the magnetic field structure; and so on. Since our knowledge of the distribution of any one of these components is still rather incomplete, modeling emission that arises from the coupling of two or more of them is a daunting task, requiring numerous assumptions to develop a complete model that can be compared with observations. 

The value of such an exercise lies in identifying the factors that have the largest influence on determining the properties of the observed emission, with the hope that eventually {\it all} of the different diffuse emission measurements can be combined to yield a more complete picture of the Galaxy.  In this paper, we present models of the stellar and diffuse infrared emission in the Galactic plane. Although investigations of the diffuse infrared emission have been carried out in the past, to our knowledge this is the first work to focus on the mid-infrared (3.6-8.0 $\mu$m) diffuse emission from polycyclic aromatic hydrocarbons (PAHs), which is much more sensitive to the distribution of the young stars than the longer wavelength dust emission probed by IRAS and COBE/DIRBE.

The large-scale distribution of stars and diffuse dust emission in the Milky-Way has been the subject of numerous studies. On one hand, powerful three-dimensional stellar distribution models have been developed to simulate the inventory of stellar populations as a function of location in the Galaxy and position on the sky. The Besan\c{c}on model \citep{robin:86:71, bienayme:87:94, robin:96:125, robin:03:523},  the SKY model \citep{wainscoat:92:111, cohen:93:1860, cohen:94:582, cohen:95:874}, and the TRILEGAL model \citep{girardi:05:895} are three of the most widely known such models aimed at reproducing the stellar populations of the Galaxy. Such models typically include various components to describe the Galaxy, such as a bulge, halo, disk(s), spiral arms, and so on, and include stars at various masses, metallicities, and stages of evolution.

On the other hand, analytic models of the diffuse dust emission in the Galaxy have also been developed to reproduce the large-scale mid- and far-infrared emission. For example, \citet{sodroski:97:173} used COBE/DIRBE data from 12$-$240\microns to determine the best-fitting abundances and temperature of large dust grains, the abundance of very small transiently heated dust grains and PAHs, and the energy density of the interstellar radiation field as a function of Galactocentric radius. Similarly, \citet{davies:97:679} fit a model for the emission from cool (18-22K) dust to the COBE/DIRBE far-infrared data. More recently, the \citet{planck:11:A21} fit the observations of the Galaxy by the Planck satellite, using a model similar to that used by \citeauthor{sodroski:97:173}

\citet{drimmel:00:L13} carried out a qualitative comparison of COBE/DIRBE K-band and 240\microns observations, and found that while the near-infrared shows evidence only for two stellar arms, the far-infrared observations are consistent with four arms (in agreement with the influential HII region study of \citealt{georgelin:70:349}), leading to the conclusion that the main Galactic potential is two-armed, but that the structure of the dust and gas responding to the potential is more complex, and can be adequately characterized as four-armed. \citet{drimmel:01:181} extended this by developing a quantitative model to fit the above data, which consisted of a parametric model of the dust distribution and associated far-infrared emission. This dust distribution was then used by \citet{drimmel:03:205} to predict the three-dimensional K-band extinction.

What has been lacking in these models however is a fully self-consistent treatment of the {\it source} of the dust heating -- the photons produced by stars of various temperatures and luminosities -- and the {\it target} of the heating -- grains of different sizes and PAHs -- distributed throughout the disk of the Galaxy. This was addressed in particular by \citet{porter:05}, who developed a self-consistent radiative transfer model of the stellar and dust emission in order to study the propagation of cosmic-rays in the Galaxy, and computed the interstellar radiation field as a function of position in the Galaxy, assuming azimuthal symmetry. The model was subsequently used and refined in \citet{moskalenko:06:L155}, \citet{porter:06:L29}, and \citet{porter:08:400}, and was successfully compared to the local integrated all-sky interstellar radiation field as seen from the Sun.

In this paper we present a fully self-consistent 3-dimensional radiative transfer model of the Galaxy at infrared wavelengths, considering both stellar and dust emission, and we compare the model to the distribution of emission on the sky from near-infrared to far-infrared wavelengths. The model is self-consistent in the sense that the heating of the dust does not follow an analytical prescription, but is instead directly computed from the stellar populations, similarly to the \citeauthor{porter:05} model. The main aims of this paper with this model are the following:

\begin{itemize}
\item To determine whether an existing Galactic model for stellar populations and dust -- here the SKY model -- used in conjunction with the dust properties determined by \citet{draine:07:810}, which include transiently heated very small grains and PAH molecules, adequately reproduces the stellar and diffuse emission seen at mid- and far-infrared wavelengths.
\item To determine whether any modifications are needed to this model to reproduce the observations, and whether any new insights on Galactic structure can be gained. However, we stress that the aim is not to provide an exhaustive parameter space study, as given the number of stellar populations and components, the number of parameters describing the model is of the order of several hundred. Rather, the aim is to determine possible and realistic modifications that improve the model.
\item To determine the relative contributions of resolved and unresolved stellar flux, emission from PAH molecules and dust grains, and scattering at infrared wavelengths.
\item To determine the relative importance of various stellar populations in heating dust grains and exciting PAH molecules.
\end{itemize}

In Section \S\ref{sec:observations} we describe the observational datasets used in this paper. In \S\ref{sec:model}, we describe the initial model used, the dust properties, and the radiative transfer code. In \S\ref{sec:results}, we show results for both the initial model (\S\ref{sec:initial}) as well as an improved model (\S\ref{sec:improved}). In \S\ref{sec:analysis} we examine the main contributors to the observed flux and to the heating of the dust (\S\ref{sec:contributions}), 
we study how much IRAC stellar flux is likely to be unresolved (\S\ref{sec:unresolved}), and we generate images of the Galaxy from an external viewpoint (\S\ref{sec:external}). Finally, in \S\ref{sec:summary} we summarize our findings.

\section{Observations}

\label{sec:observations}

In this paper we make use of the \textit{Spitzer} GLIMPSE \citep{benjamin:03:953, churchwell:09:213} and MIPSGAL \citep{carey:09:76} surveys of the Galactic mid-plane,
specifically GLIMPSE/MIPSGAL~I, which cover $10$\degrees $\le|\ell|\le65$\degrees and $|b|\le1$\degrees, and GLIMPSE/MIPSGAL~II, which fill in the region for $|\ell|<10$\degrees, with $|b|\le1$\degrees for $|\ell|>5$\degrees , $|b|\le1.5$\degrees for 2\degrees$<|\ell|\le5$\degrees, and $|b|\le2$\degrees for $|\ell|\le2$\degrees. The total area covered by these surveys is 274\sqdeg.

The GLIMPSE surveys were carried out using the InfraRed Array Camera \citep[IRAC;][]{fazio:04:10} at 3.6, 4.5, 5.8, and 8.0\microns. The MIPSGAL surveys were carried out using the Multiband Imaging Photometer for Spitzer \citep[MIPS;][]{rieke:04:25} at 24 and 70\microns, though only the processed 24\microns data are available and are included in this study.

In addition to the \textit{Spitzer} data, we include data from the Improved Reprocessing of the IRAS Survey \citep[IRIS;][]{miville-deschenes:05:302} at 60 and 100\microns to provide further constraints on the dust emission.

At the different wavelengths considered, the total emission in a given direction consists of stellar emission resolved into point sources, unresolved stellar emission, and diffuse interstellar emission. The relative contribution of these three types of emission change both as a function of wavelength (diffuse emission increases with wavelength), and position in the Galaxy (unresolved stellar emission decreases with increasing angle from Galactic center). Here, longitude and latitude profiles were constructed for the four IRAC bands, the MIPS 24\microns data, and the IRIS data, by resampling the full data into 3' by 3' bins, including the resolved stellar flux. The re-binned data were then collapsed into selected longitude and latitude profiles to compare the observations to the radiative transfer model presented in \S\ref{sec:results}. In this paper, we only use the rectangular region for $|\ell| < 65^\circ$ and $|b|<1^\circ$ to facilitate computing longitude and latitude profiles.

\section{Model}

\label{sec:model}

Our self-consistent radiative transfer model of the stellar and dust emission at infrared wavelengths consists of two key ingredients: a model for the distribution of the stellar populations, and a model for the characteristics and distribution of the dust. The initial distributions used are described in \S\ref{sec:populations} and in \S\ref{sec:dust_distribution}, but these are then modified in \S\ref{sec:improved} to provide a better fit to the observations. The dust properties assumed are described in \S\ref{sec:dust_properties}. Finally, the radiative transfer model is described in \S\ref{sec:rt}.

\subsection{Stellar populations}

\label{sec:populations}

To model the distribution of various stellar populations, we implemented the SKY model from \citet[hereafter W92]{wainscoat:92:111} and \citet{cohen:93:1860, cohen:94:582, cohen:95:874}. The choice of the SKY model over more recent models such as the  Besan\c{c}on or TRILEGAL models is that the former provides a prescription for spiral arms, which are essential for the present work, since we expect the diffuse PAH emission at IRAC wavelengths to depend on the distribution of massive stars, which are concentrated in the spiral arms. In addition, the three-dimensional distribution of the stellar populations in the SKY model is separated into spectral classes, making it straightforward to represent in a radiative transfer model.

The SKY model includes five components to describe the Galaxy: an exponential disk, a bulge, a halo, spiral arms, and a ring. We do not include the halo: as it is defined in W92 by a projected rather than a spatial distribution, it is more difficult to include in the radiative transfer model. However, it has a very low normalization factor relative to the disk (1:1250; \citealt{cohen:95:874}) and would have no noticeable impact on the \textit{Spitzer} or IRAS observations in the Galactic mid-plane. The remaining four main components are briefly described in the following sections.

The model includes 87 different types of Galactic sources, including pre-main-sequence stars, main-sequence stars, giants, asymptotic giant branch (AGB) stars, supergiants, and planetary nebulae. Each component includes a normalization factor ($\rho_D$, $\rho_B$, $\rho_A$, $\rho_r$), and in some cases other parameters that depend on the source type, denoted $S$. For example, O, B, and A stars are absent from the bulge (typically composed of more evolved stars), while lower-mass main sequence stars are absent from the spiral arms (typically composed of relatively young stars). In the following sections, all parameter values that are functions of the source type $S$ are given in Table~2 of W92. The following sections give an overview of the distributions used; the reader is referred to W92 for more details. We note at the outset that the parameters characterizing these components, and in some cases even their existence (such as the molecular ring), are uncertain. In the discussion that follows, we will explore the effect of changing some of these components. 

\subsubsection{The exponential disk}

The exponential disk has a density distribution given in cylindrical polar coordinates by
$$
\rho(R, z, S) = \rho_D(S)\exp{\left[-\frac{R-R_0}{h} - \frac{|z|}{h_z(S)}\right]}
$$
where $h_z(S)$ is the disk scale height, which depends on the spectral class $S$, $R_0=8.5$\kpc is the distance from the Sun to the Galactic center, and $h=3.5$\kpc is the radial scale-length. The exponential disk is truncated at $R_{\rm max}=15$\,kpc.

\subsubsection{The bulge}

The bulge has a density distribution given by
$$
\rho(R, z, S) = \rho_B(S)\,x^{-1.8}\exp{\left(-x^3\right)}.
$$
where
$$
x = \frac{\sqrt{R^2+k_1^2\,z^2}}{R_1}
$$
In the W92 model, the bulge axis ratio $k_1$ is set to 1.6, and the bulge radius $R_1$ is set to 2\,kpc.

\subsubsection{The spiral arms}

\begin{table}
\centering
\caption{Spiral Arm Parameters \label{tab:arms}}
\begin{tabular}{lccccc}
\hline\hline
Arm & $\alpha$ & $R_{\rm max}$ & $\theta_{\rm min}$ & extent & width \\
 & & (kpc) & (rad) & (rad) & (kpc) \\
\hline
1  & 4.25 & 3.480  & 0.000 & 6.00  & 0.75 \\
1' & 4.25 & 3.480  & 3.141 & 6.00 & 0.75 \\
2  & 4.89 & 4.900 & 2.525 & 6.00 & 0.75 \\
2' & 4.89 & 4.900 & 5.666 & 6.00  & 0.75 \\
L  & 4.57 & 8.100  & 5.847 & 0.55 & 0.30 \\
L' & 4.57 & 7.591 & 5.847 & 0.55 & 0.30 \\
\hline
\end{tabular}
\end{table}

The original W92 model included four main logarithmic spiral arms, and one spur in the Solar neighborhood. \citet{cohen:94:582} subsequently replaced the single local spur by two narrower spurs on either side of the Sun. 
The shape of the arms is parameterized as
$$
\theta(R) = \alpha\,\log{\left(\frac{R}{R_{\rm min}}\right)} + \theta_{\rm min},
$$
where $\alpha$ is the winding constant, $R_{\rm min}$ is the inner radius, and $\theta_{\rm min}$ is the angle at the inner radius. The arms go out to $R_{\rm max}$, but are truncated above a certain length. The parameters for the arms (compiled from Table 2 of W92 and \citealt{cohen:94:582}) are given in Table~\ref{tab:arms}. The arms follow the same radial and vertical exponential distributions as the disk, and have a constant width. Thus, their density distribution is effectively given by
$$
\rho(R, z, S) = \rho_A(S)\exp{\left[-\frac{R-R_0}{h} - \frac{|z|}{h_z(S)}\right]}
$$
inside the arms, and $\rho(R,z,S)=0$ outside. As mentioned previously, the stellar population in this component is dominated by young, massive stars. 

\subsubsection{The stellar (molecular) ring}

The SKY model contains a stellar ring with a density given by
$$
\rho(R, z, S) = \rho_r(S)\exp{\left[\frac{-\left(R-R_r\right)^2}{2\sigma_r^2}\right]}\exp{\left[- \frac{|z|}{h_z(S)}\right]}
$$
where $R_r=6.75$\kpc is the ring radius, and $\sigma_r=0.96$\kpc is the ring width. Since this component was motivated by an annulus of molecular gas in the inner galaxy, the stellar populations in this component are also weighted to young massive stars, as is the case for the spiral arms. 

\subsection{Dust distribution}

\label{sec:dust_distribution}

The second main ingredient for the Galactic model is a distribution of dust, which is initially taken from the extinction distribution of W92. For the dust properties described in \S\ref{sec:dust_properties}, the extinction distribution from W92 corresponds to
\begin{equation}
\label{eq:dust}
\rho(R, z, S) = \rho_d \exp{\left[-\frac{R}{h_d} - \frac{|z|}{z_d}\right]}
\end{equation}
where $h_d=3.5$\,kpc, $z_d=100$\,pc, and $\rho_d=10^{-25}$\,g/cm$^3$.

Given the clumpiness of the molecular gas of the Galaxy, the smooth exponential distribution of dust is clearly an oversimplification, but serves as a useful starting point for the model presented here. 

\begin{table*}
\centering
\caption{Variance between the models and the observations\label{tab:variance}}
\begin{tabular}{lcccccccc}
\hline\hline
 & \multicolumn{2}{c}{Original (\S\ref{sec:initial})} & \multicolumn{2}{c}{Two spiral arms (\S\ref{sec:twoarm})} & \multicolumn{2}{c}{Inner dust hole (\S\ref{sec:hole})} & \multicolumn{2}{c}{Increased PAHs (\S\ref{sec:morepah})} \\
Wavelength & Longitude & Latitude & Longitude & Latitude & Longitude & Latitude & Longitude & Latitude \\
\hline
3.6\,$\mu$m & 0.0133 & 0.0056 & 0.0071 & 0.0019 & 0.0057 & 0.0008 & 0.0064 & 0.0047 \\
4.5\,$\mu$m & 0.0248 & 0.0097 & 0.0179 & 0.0062 & 0.0155 & 0.0009 & 0.0113 & 0.0002 \\
5.8\,$\mu$m & 0.0756 & 0.0246 & 0.0235 & 0.0046 & 0.0255 & 0.0093 & 0.0112 & 0.0011 \\
8.0\,$\mu$m & 0.1147 & 0.0542 & 0.0374 & 0.0142 & 0.0444 & 0.0242 & 0.0187 & 0.0016 \\
24\,$\mu$m  & 0.0643 & 0.0095 & 0.0331 & 0.0041 & 0.0323 & 0.0027 & 0.0406 & 0.0074 \\
60\,$\mu$m  & 0.1191 & 0.0560 & 0.1640 & 0.1027 & 0.1163 & 0.0331 & 0.1061 & 0.0251 \\
100\,$\mu$m & 0.0580 & 0.0260 & 0.0739 & 0.0527 & 0.0436 & 0.0071 & 0.0401 & 0.0049 \\
\hline
Total       & 0.0671 & 0.0265 & 0.0510 & 0.0267 & 0.0405 & 0.0112 & 0.0335 & 0.0064 \\
\hline
\end{tabular}
\end{table*}

\subsection{Dust properties}

\label{sec:dust_properties}

The dust model adopted is from \citet{draine:07:810}. The model consists of a mixture of carbonaceous and amorphous silicate grains with a size distribution from \citet{weingartner:01:296} that reproduces interstellar extinction curves with R$_V$=3.1. The model uses the renormalization relative to H from \citet{draine:03:241}. The size distribution includes a population of small grains in two log-normal components in the size distribution, to represent transiently heated very small grains and PAH molecules. The model from \citeauthor{draine:07:810} used here is the one with a mass fraction of PAHs set to q$_{\rm PAH}$=4.6\%.

\citeauthor{draine:07:810} computed emissivities for dust grains illuminated by the interstellar radiation field in the solar neighborhood from \citet{mathis:83:212} scaled by a factor -- denoted $U$ -- which varied between 0.5 and $10^7$. As described in \citet{draine:07:810}, the ionization fraction of the PAHs is a function of grain size. The emissivities and opacities used were identical to those publicly available\footnote{\texttt{http://www.astro.princeton.edu/~draine/dust/irem.html}} but were separated into three size regimes (B. Draine, private communication): ultra small grains ($a<20$\AA), very small grains (VSGs; $20$\AA$<a<200$\AA), and big grains ($a>200$\AA). These three regimes contain 5.86\%, 13.51\%, and 80.63\% of the mass respectively in the dust model. Since the ultra-small grains are dominated by PAH molecules, we will refer to them as PAHs in the remainder of this paper (but we note that they are not strictly equivalent).

\subsection{Radiative transfer model}

\label{sec:rt}

\subsubsection{Model set-up}

The radiative transfer models presented in this paper were computed with the Monte-Carlo three-dimensional radiative transfer code {\sc Hyperion}\footnote{\texttt{http://www.hyperion-rt.org}}  \citep{robitaille:11:A79}. This code can read in any arbitrary dust geometry, with multiple dust components, and multiple sources, making it perfectly suited to this project. It also allows images to be produced relative to an observer inside the density grid, which we make use of in this paper. The code implicitly solves the full equation of radiative transfer, including optical depth effects, scattering, polarization, and emission under the assumption of radiative equilibrium.

The stellar and dust densities were discretized onto a cylindrical polar grid with 200 radial, 50 vertical, and 100 azimuthal cells. The radial cells were distributed linearly between 0 and 15\,kpc. The vertical cells were logarithmically spaced from 1\,pc to 3\,kpc above and below the mid-plane, with the region inside $\pm 1$\,pc divided into two cells. The azimuthal cells were uniformly distributed from 0 to 2$\pi$.

Rather than treating sources individually, which would be unfeasible computationally, `diffuse' sources of emission were used to represent populations of stars: for each spectral class, each cell in the grid was given a probability for emission, and any photon emitted from that source was given the spectrum corresponding to the spectral class. Each diffuse source was given a total luminosity derived from the total number of `real' sources.

By default, the calibrated spectra used for each spectral class were derived from the absolute magnitudes given in Table 2 of W92 for BVJHK and 12\microns \& 25\microns. However, for main-sequence stars, giant, and supergiants, we used \citet{allen} to transform the spectral types into effective temperatures and surface gravities, which we subsequently used to interpolate stellar photosphere models from \citet{castelli:04}. These model spectra were then scaled to the absolute magnitudes given in Table~2 of W92. The motivation for doing this is that the B magnitudes given in the SKY model are not sufficient to properly characterize the UV portion of the spectrum, which is crucial for the heating of the PAHs and VSGs. For the `Young OB' source type in W92, which is meant to represent OB associations, we fit the \citeauthor{castelli:04} spectra to the magnitudes supplied by W92, and found that an effective temperature of 15,000\,K provided the best fit to all wavelengths.

Each of the dust types (PAHs, VSGs, and big grains) was given the spatial distribution from Equation (\ref{eq:dust}), or the modified spatial distribution defined in \S\ref{sec:hole}, scaled by the mass fraction of the dust types (c.f. \S\ref{sec:dust_properties}).

\subsubsection{Caveats}

\label{sec:caveats}

The \textsc{Hyperion} radiative transfer code uses pre-computed emissivities for the grains for a range of interstellar radiation field strengths, rather than computing the transient heating of very small grains and PAH molecules exactly. As described in \S\ref{sec:dust_properties}, the dust emissivities were pre-computed for a fixed illuminating spectral shape -- but variable intensity $U$ -- so for regions in the Galaxy with different radiation field shapes, the shape of the emissivity will likely be different.  The primary effect of a different spectral shape is to change the amount of radiation absorbed by and therefore heating the PAHs. Since PAHs have an opacity that is highest in the UV, harder radiation fields with a fixed intensity $U$ will lead to higher excitation levels. This can be accounted for by quantifying the radiation field not by $U$ (which ignores the wavelength dependence of the emission) but by the power of the radiation field absorbed by the grains (per unit mass):
$$
\dot{A} = \int{4\pi J_{\rm isrf}\kappa_\nu d\nu}
$$
where $J_{\rm isrf}$ is the mean intensity of the radiation field. Thus, two radiation fields with the same intensity $U$ but different spectral shapes -- which would excite the PAHs by different amounts -- would have correspondingly different values of $\dot{A}$. Therefore, selecting the emissivities based on $\dot{A}$ is a better approximation than simply using $U$ and takes into account the effects of spectral shape to first order.

With this effect taken into account, the difference between the emissivities due only to the change in spectral shape for a constant $\dot{A}$ is quantitatively small for our purposes. As shown in Figure 7b of \citet{draine_pah_review}, for two very different radiation fields with the same\footnote{As described in Section 6 of \citet{draine_pah_review}, the 20,000\,K blackbody is adjusted in intensity to give the same power per H absorbed by dust, which is equivalent to saying with our definition that the two radiation fields have the same $\dot{A}$} $\dot{A}$, the difference in the PAH emissivities is less than a factor of 1.5. In this case the emissivities are systematically offset, and the colors do not change noticeably. Since at least some of the ISM will in fact have an interstellar radiation field close to that used to compute the dust emissivities, this effect may be smaller in reality.

As shown in see Figure 7a of \citet{draine_pah_review}, the difference in the emissivities between fully ionized and fully neutral PAHs is more important. In particular, different levels of ionization can lead to differences in emissivities of factors of two, as well as significant changes in colors.

We conclude that the assumption of the specific dust model used, which includes an assumption for the ionization level of the PAHs, is the main caveat in the radiative transfer models, while the method of choosing the emissivities based on $\dot{A}$, while approximate, appears to be adequate for our purposes.

\rm

\subsubsection{Model output}

Model images were computed for the survey area and collapsed into longitude and latitude profiles in the same way as the data. The latitude resolution of the profiles is 3', while the longitude resolution is 1$^\circ$. The images were computed for 160 wavelength bins logarithmically spaced from 3 to 140 microns, and were subsequently convolved with the transmission curves for IRAC, MIPS, and IRAS (following Appendix A of \citealt{robitaille:07:328}). The convolution is especially important because the PAH features in the emissivities cause the flux to vary rapidly with wavelength in the mid-infrared, and picking a single wavelength instead of taking into account the proper transmission curve could result in the model fluxes being wrong by a factor of two or more.

For all four models presented in Section \ref{sec:results} we have computed the variance between the data and the model in log$_{10}$ space to quantify the goodness of fit, and we list these in Table~\ref{tab:variance}. We do not compute $\chi^2$ values, since the model is clearly misspecified -- that is, the model is only a rough approximation of reality because it does not take into account the small-scale features in the emission (which we are not attempting to reproduce) -- so that absolute $\chi^2$ values and associated likelihoods would not be meaningful. We compute the variances separately for longitude and latitude profiles, and provide both the variances for individual wavelengths, and the overall values.

\section{Results}

\label{sec:results}

\subsection{Original model}

\label{sec:initial}

In this section, we show the results from the model as described in \S\ref{sec:model} with no modifications.

\begin{figure*}
\begin{center}
\includegraphics[width=0.49\textwidth]{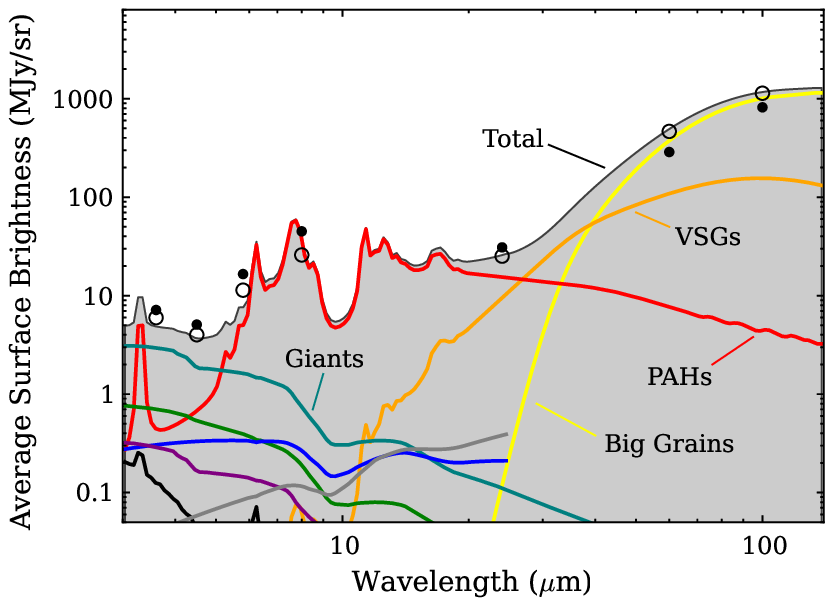}
\includegraphics[width=0.49\textwidth]{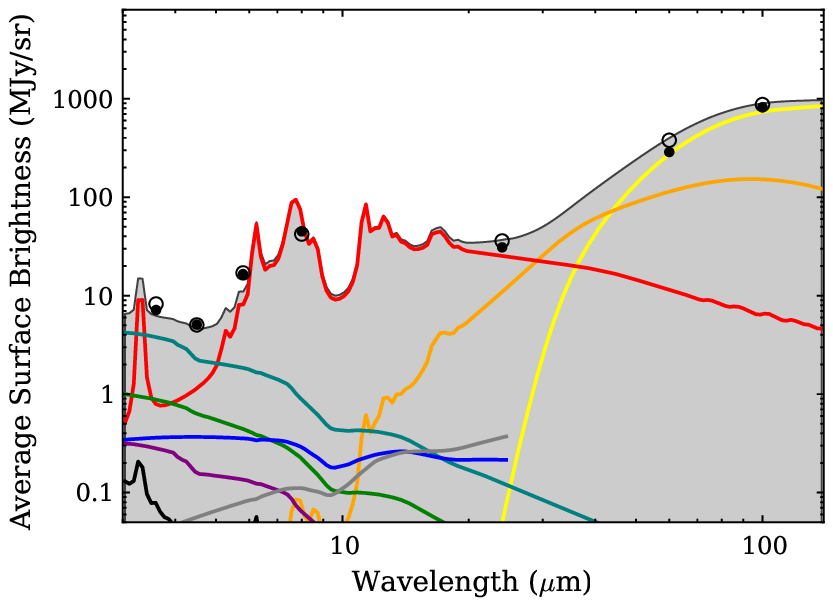}
\caption{Spectrum of the average surface brightness inside $-65^\circ < \ell < 65^\circ$ and $-1^\circ < b < 1^\circ$ for the initial model (left) and the final modified model (right). The thick colored solid lines represent different components of the model: supergiants (purple), AGB stars (blue), giants (teal), main sequence stars (green), other stellar types (dark gray), dust emission from PAHs (red), VSGs (orange), and big dust grains (yellow), and scattered light (black). The major components are labelled in the left panel. The shaded gray area shows the total flux. The open black circles show the model fluxes after convolution with the transmission curves, and the filled black circles show the observations.\label{fig:spectra}}
\end{center}
\end{figure*}

\begin{figure*}
\begin{center}
\includegraphics[width=0.98\textwidth]{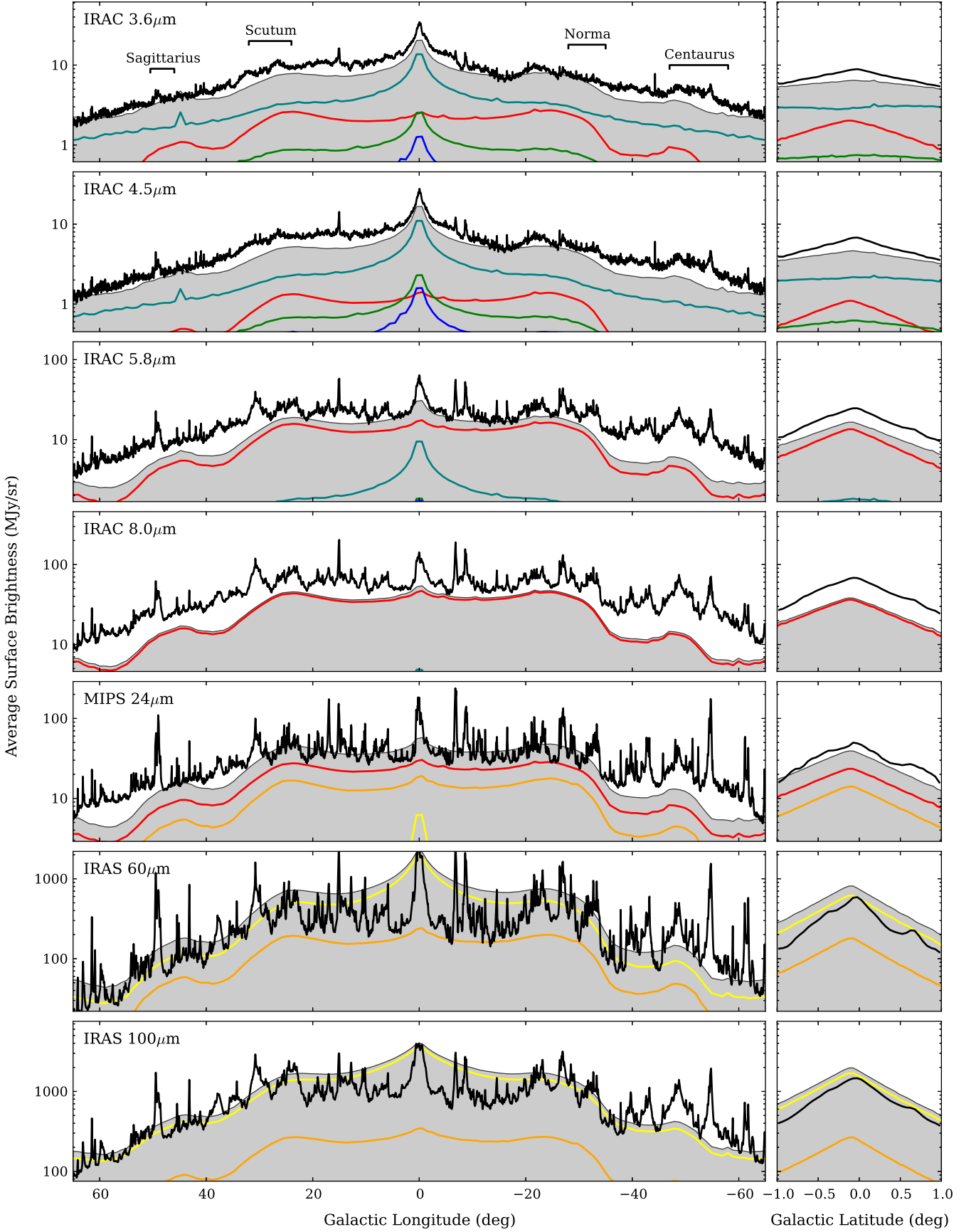}
\caption{The longitude and latitude surface brightness profiles in the range $-65^\circ < \ell < 65^\circ$ and $-1^\circ < b < 1^\circ$ for the observations and for the initial model (\S\ref{sec:initial}). The thick black lines show the observed surface brightness, while the colored lines and the gray shaded area show the model surface brightness, with the same colors as used in Figure~\ref{fig:spectra}. The `traditional' spiral arm tangencies are indicated in the top panel.\label{fig:initial_profiles}}
\label{default}
\end{center}
\end{figure*}

\begin{figure*}
\begin{center}
\includegraphics[width=0.98\textwidth]{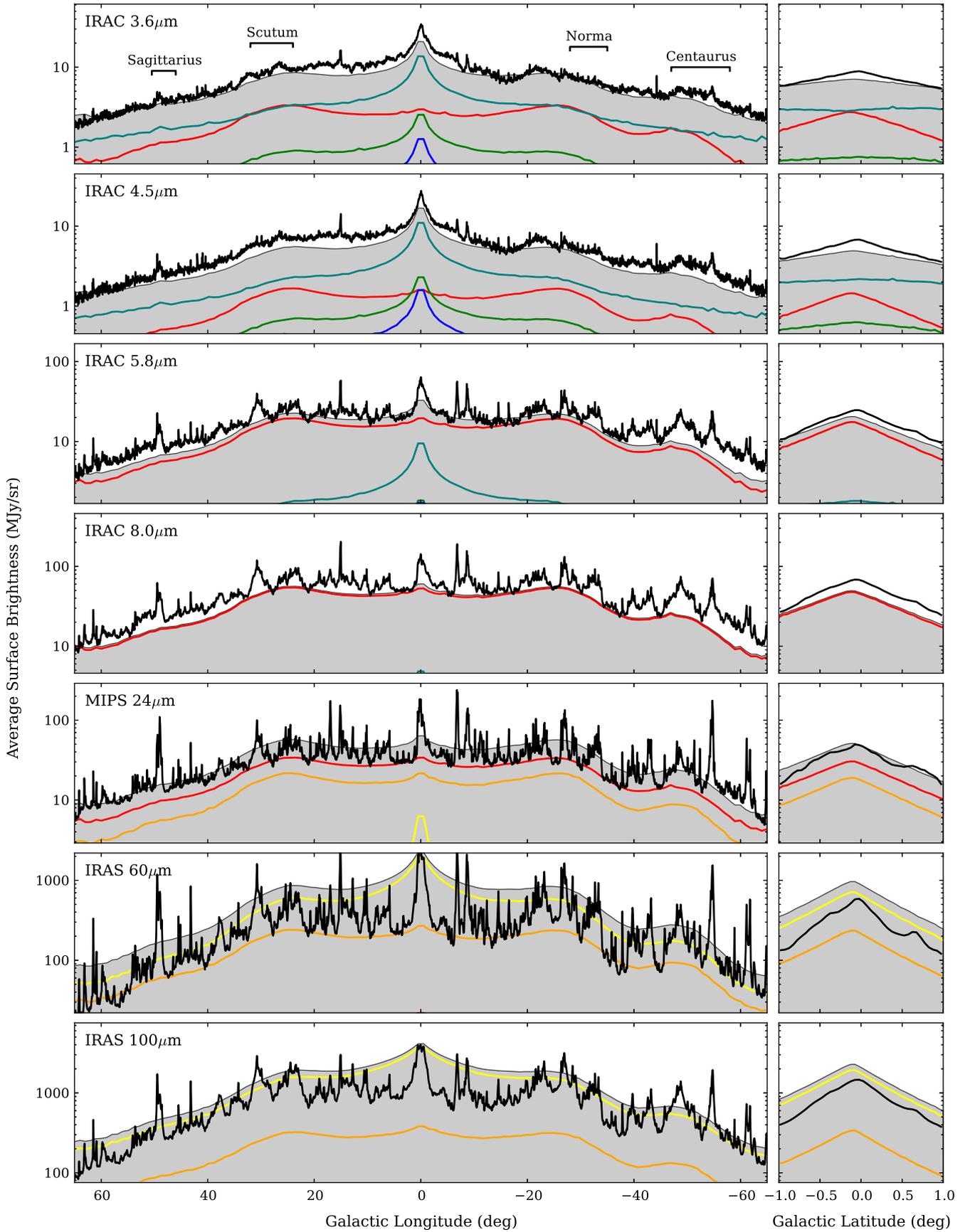}
\caption{The longitude and latitude surface brightness profiles as for Figure~\ref{fig:initial_profiles}, but for the model including two major and two minor spiral arms instead of four main spiral arms (\S\ref{sec:twoarm}).\label{fig:twoarm_profiles}}
\label{default}
\end{center}
\end{figure*}

\begin{figure*}
\begin{center}
\includegraphics[width=0.98\textwidth]{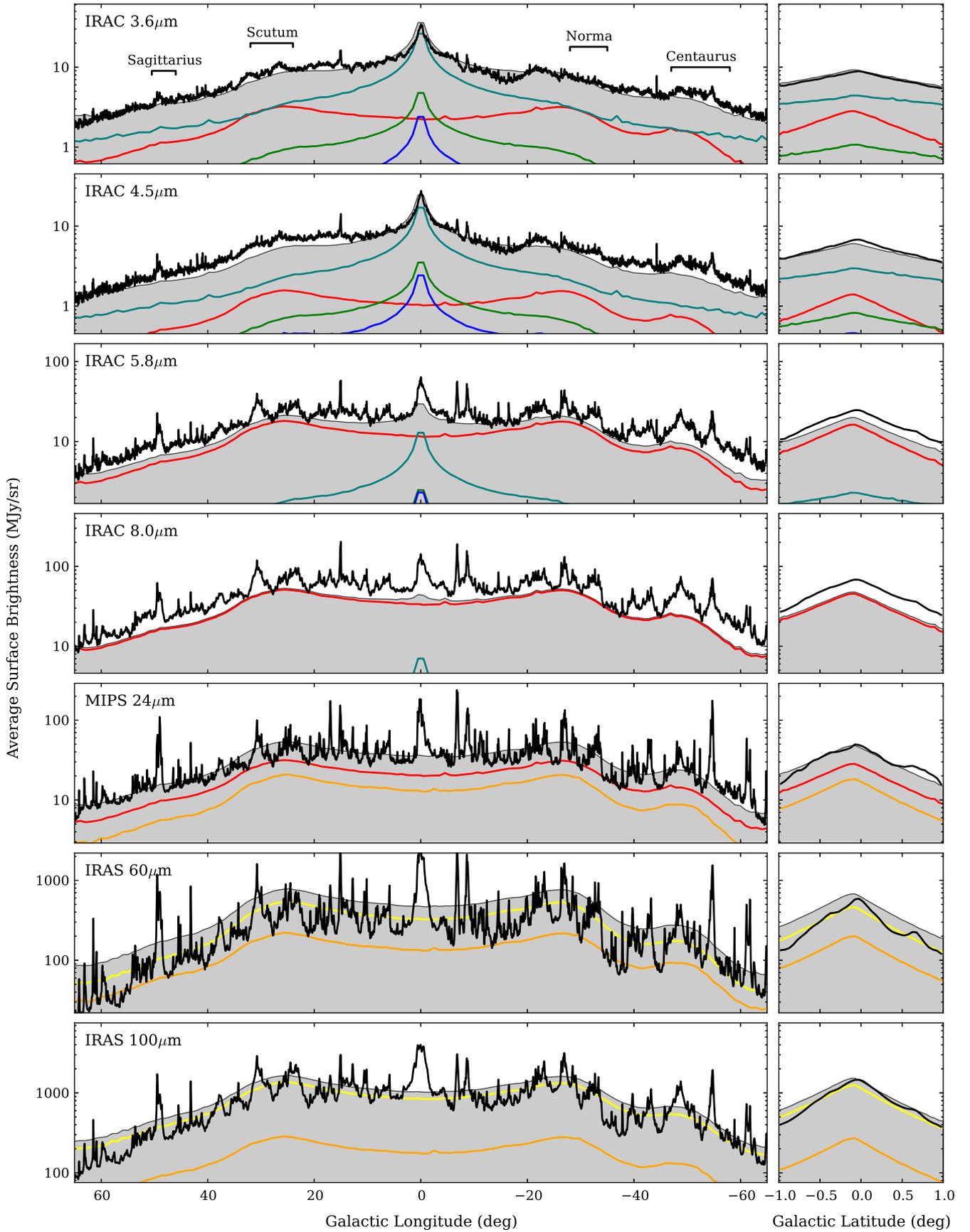}
\caption{The longitude and latitude surface brightness profiles as for Figure~\ref{fig:twoarm_profiles}, with a deficit in the dust distribution in the central few kpc of the Galaxy (\S\ref{sec:hole}).\label{fig:hole_profiles}}
\label{default}
\end{center}
\end{figure*}

\begin{figure*}
\begin{center}
\includegraphics[width=0.98\textwidth]{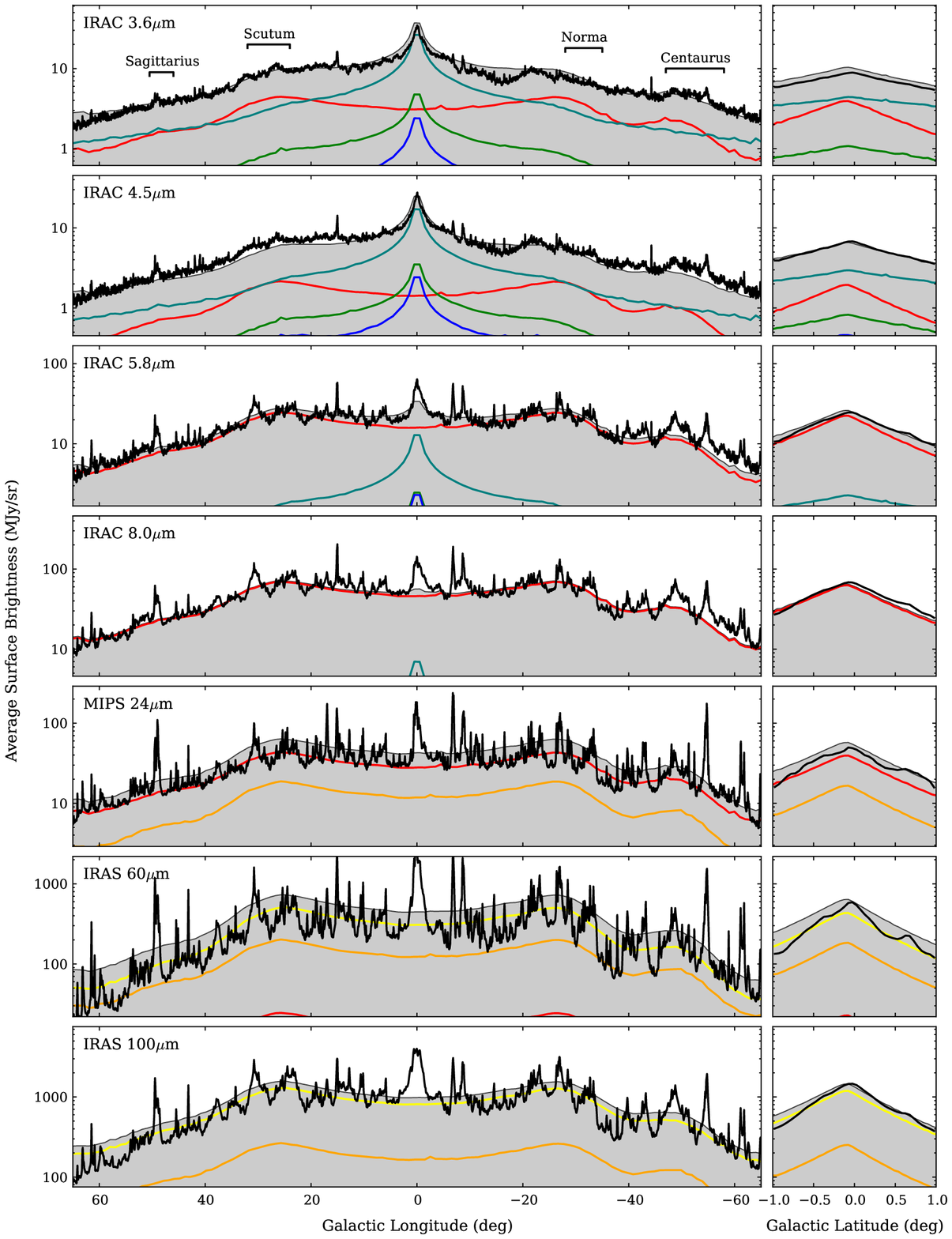}
\caption{The longitude and latitude surface brightness profiles as for Figure~\ref{fig:hole_profiles}, with 50\% more PAHs (\S\ref{sec:morepah}).\label{fig:morepah_profiles}}
\label{default}
\end{center}
\end{figure*}

Figure~\ref{fig:spectra} shows the average spectrum of the surface brightness inside $-65^\circ < \ell < 65^\circ$ and $-1^\circ < b < 1^\circ$ for both the observations and the model. The model underestimates the surface brightness at IRAC wavelengths and overestimates it in the far infrared at IRAS wavelengths.

In addition to the averaged surface brightness spectrum across the survey area, we also examine the spatial distribution of the emission. Figure~\ref{fig:initial_profiles} shows the observed and model longitude and latitude profiles from 3.6 to 100\microns. The emission in the model is of the right order of magnitude, and reproduces a few important features of the observations, including the scale height of the emission (longwards of IRAC 5.8\microns) and the flattening of the IRAC 5.8 and 8.0\microns emission inside $|\ell|<30^\circ$, but there are also a number of issues:

\begin{itemize}

\item the model surface brightness is too low by a factor of two or more for $\ell < -30^\circ$ for IRAC 3.6\microns to MIPS 24\microns, and for $\ell > 30^\circ$ for IRAC 5.8\microns to MIPS 24\microns. Since $\ell=30^\circ$ corresponds to the tangency of the ring component for our model, this suggests that the model for the disk and/or spiral structure outside this radius are not adequate.

\item the model surface brightness is too high for $|\ell| < 30^\circ$ for IRAS 60\microns and 100\microns, which suggests that our model contains too much dust inside $R=4$ kpc. It has previously been suggested by numerous authors that a hole exists in the distribution of gas within a few kpc of the Galactic center \citep[c.f.][]{ortiz:93:90} -- presumably due to the bar -- which would explain the excess flux predicted by the model here.

\item the IRAC 3.6 and 4.5\microns latitude surface brightness profiles are not as peaked as the data. This suggests that either the scale height of the stellar populations is not small enough, or that dust extinction is too large in the mid-plane, since dust extinction is included in the model. The shape of the latitude profiles is dominated by the inner Galaxy ($|\ell|<30^\circ$) since that is where most of the emission originates, so changes in this range will have the largest impact on the latitude profiles.

\end{itemize}

\subsection{Improved model}

\label{sec:improved}

In this section, we examine how the model presented in \S\ref{sec:initial} changes as we modify our initial assumptions in ways that have been motivated by previous investigations of Galactic structure.  The modifications explored are changes to the parameters for the spiral arms (\S\ref{sec:twoarm} and Figure~\ref{fig:twoarm_profiles}), modifications of the dust distribution interior to $R=4$ kpc (\S\ref{sec:hole} and Figure~\ref{fig:hole_profiles}), and modifications to the abundance of PAHs (\S\ref{sec:morepah} and Figure~\ref{fig:morepah_profiles}).  We also explored the effect of concentrating the dust in the arms, but found that the simulations most consistent with the observations were the ones in which the dust is substantially broader than the stellar arms.

\subsubsection{Modified spiral structure}

\label{sec:twoarm}

Our modifications to the spiral structure model are motivated by the low model surface brightnesses for $|\ell|>30^\circ$ at IRAC and MIPS wavelengths relative to the observations

Previous research has shown that when using stellar tracers in the near- or mid-infrared, there is only evidence for two of the spiral arms in the Galaxy \citep[e.g.][]{drimmel:00:L13, drimmel:01:181}, specifically the Scutum-Centaurus arm and the Perseus arm. On the other hand, surveys in the far-infrared and radio show that models with four arms are a better match. This suggests that the main potential of the Galaxy is in fact two armed, but that the non-linear response of the gas and dust flowing through the stellar potential is more complex, and includes arms, or large spurs, between the two major spiral arms.

We therefore modify the model so that the Scutum-Centaurus and Perseus arms (2 and 2' in Table~\ref{tab:arms}) become the dominant arms by increasing the normalization factor of the O and B spectral class populations in these arms by a factor of two. Rather than completely eliminating the Norma and Sagittarius arms (1 and 1' in Table~\ref{tab:arms}), which does not provide a good fit to the dust-dominated bands, we eliminate all stars except those belonging to the O and B spectral classes, which we leave unchanged, under the assumption that stars are indeed forming in these secondary arms but do not stay there in the long term.

The shape of these four arms is also changed to Gaussian profiles in the radial direction, with $\sigma=550$\,pc. Finally, we modify the spiral arm parameters slightly to provide a better match to the shape of the longitude profiles. The updated spiral arm parameters are listed in Table~\ref{tab:updated_arms}. The local spurs are left unchanged. The resulting model longitude and latitude profiles, shown in Figure~\ref{fig:twoarm_profiles} provide a better fit to the observations at IRAC and MIPS wavelengths for $|\ell|>30^\circ$, although we note that the IRAC 5.8 and 8.0\microns fluxes are still slightly too low in that longitude range. This is reflected in the decrease in the variance between the model and the data listed in Table~\ref{tab:variance} for IRAC and MIPS wavelengths.

\begin{table}
\centering
\caption{Updated Spiral Arm Parameters \label{tab:updated_arms}}
\begin{tabular}{lcccccc}
\hline\hline
Arm & $\alpha$ & $R_{\rm max}$ & $\theta_{\rm min}$ & extent & $\sigma$ & width \\
 & & (kpc) & (rad) & (rad) & (kpc) & (kpc) \\
\hline
1  & 4.18 & 3.800 & 0.234 & 6.00 & 0.55 & - \\
1' & 4.18 & 3.800 & 3.376 & 6.00 & 0.55 & -  \\
2  & 4.19 & 4.500 & 5.425 & 6.00 & 0.55 & -  \\
2' & 4.19 & 4.500 & 2.283 & 6.00 & 0.55 & -  \\
L  & 4.57 & 8.100 & 5.847 & 0.55 & -  & 0.30 \\
L' & 4.57 & 7.591 & 5.847 & 0.55 & -  & 0.30 \\
\hline
\end{tabular}
\end{table}

\subsubsection{Modified dust distribution}

\label{sec:hole}

\begin{figure}
\begin{center}
\includegraphics[width=0.5\textwidth]{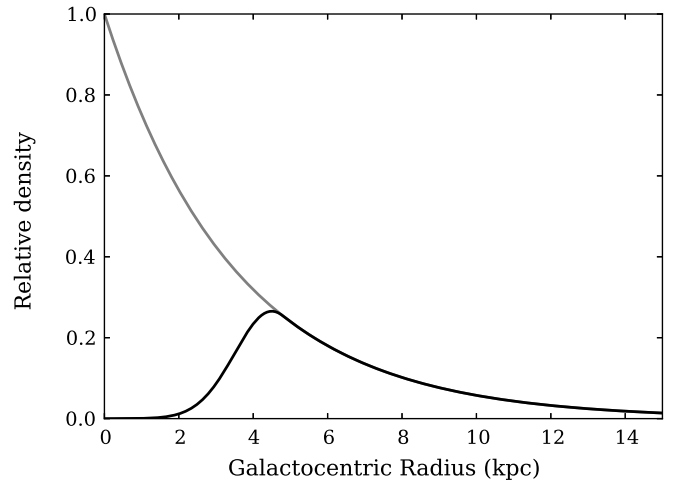}
\caption{The relative radial dependence of the dust density for the original model presented in \S\ref{sec:initial} (gray) and the model with an inner dust hole from \S\ref{sec:hole} (black).\label{fig:radial_density}}
\label{default}
\end{center}
\end{figure}

In order to provide a better fit to the IRAS 60 and 100\microns longitude profiles inside $|\ell| < 30^\circ$, we also experimented with including a hole in the dust within a few kpc of the Galactic center by modifying the radial dependence of the dust distribution from a simple exponential to a function with the following form:
$$
f_R(R) = \left\{\begin{array}{ll} f_0\,\exp{\left[-\left(R-\mu_0\right)^2/2\sigma_0^2\right]} & R < R_{\rm smooth}\\\exp{\left[-R/h\right]} & R \ge R_{\rm smooth}\end{array}\right.
$$
where $\mu_0$ and $\sigma_0$ are parameters, and $f_0$ and $R_{\rm smooth}$ are defined as the normalization constant and transition radii for which the two functions transition smoothly, meaning that the functions and their derivatives are equal. These values are:
$$
R_{\rm smooth} = \sigma_0^2 / h + \mu_0
$$
and
$$
f_0 = \frac{\exp{\left[-R_{\rm smooth} / h\right]}}{\exp{\left[-\left(R_{\rm smooth}-\mu_0\right)^2/2\sigma_0^2\right]}}
$$
We found that values of $\mu_0=4.5$\,kpc and $\sigma_0=1$\,kpc provided a model that fit the IRAS profiles well, while maintaining the fit to the IRAC and MIPS profiles. A comparison of the new density profile with the original one is shown in Figure~\ref{fig:radial_density}. The surface brightness profiles are shown in Figure~\ref{fig:hole_profiles}, and the corresponding average spectrum is shown in Figure~\ref{fig:spectra}. The MIPS 24\microns, IRAS 60\microns, and IRAC 100\microns now appear to fit the observed profiles well, while the IRAC 5.8\microns and 8.0\microns remain systematically low.

As shown in Figure~\ref{fig:hole_profiles} and Table~\ref{tab:variance}, the addition of the hole in the dust distribution also improves the fit of the IRAC 3.6 and 4.5\microns latitude profiles, which are now peaked in the same way as the observations, thanks to the reduced mid-plane extinction towards the Galactic center.

\subsubsection{Modified PAH abundance}

\label{sec:morepah}

The fact that the systematic offset at IRAC wavelengths is seen at all longitudes suggests that it cannot be solved by changes in the geometry alone, but rather by a change in the large-scale properties of the dust or radiation field. For example, one solution might be to increase the abundance of PAHs relative to larger grains. To test this, we computed models with an increased dust density for the PAHs (strictly speaking the dust grains with $a<20$\AA), and found that a model with 50\% more PAHs (corresponding to $q_{\rm PAH}\sim6.7$\%) provides a good fit to the observations\footnote{In reality, one should not increase the density of all grains with $a<20$\AA, but instead, the q$_{\rm PAH}$ fraction, which causes a bump in the grain size distribution, should be increased (since the 20\AA\,threshold is arbitrary). }. The resulting integrated spectrum is shown in Figure~\ref{fig:spectra} and the profiles are shown in Figure~\ref{fig:morepah_profiles}: the fit to the IRAC 5.8\microns and 8.0\microns profiles is significantly improved (as seen in the decrease in the variance for these bands in Table~\ref{tab:variance}), and the model now provides a reasonable fit at all wavelengths. This result suggests that -- if the model for the stellar populations is correct -- the fraction of PAHs in the ISM may be higher than previously found, although we caution that due to the large number of parameters involved, and given the caveats of the model (\S\ref{sec:caveats}), this result is only tentative.

We note that the original value of $q_{\rm PAH}=4.6$\% found by \citet{draine:07:810} was also determined by using emission from the Galactic plane, including data from the GLIMPSE survey. \citeauthor{draine:07:810} implicitly assume that the emission and colors of the GLIMPSE emission are due to optically thin emission, but our model indicates that in the mid-plane, extinction can affect the IRAC fluxes by up to 50\%. We ran a model without increasing the PAH abundance, instead computing the profiles as if the Galaxy was optically thin, and found that the profiles and total flux were a better match to the observations, and fit almost as well as when increasing the PAH abundance. Therefore, it could be that our result is consistent with \citeauthor{draine:07:810}, and that the true PAH fraction is higher than $q_{\rm PAH}=4.6$\%, but appears to be lower due to the extinction of the IRAC wavelengths relative to the far-infrared.

We also note that there may be other solutions to the systematic offset at IRAC wavelengths. For example, the clumpy nature of the ISM may affect the average energy input of the various dust types in a systematic way, resulting in different relative contributions of emission from the different dust types at different wavelengths, but we defer such an investigation to a future study.

\begin{figure*}
\begin{center}
\includegraphics[width=\textwidth]{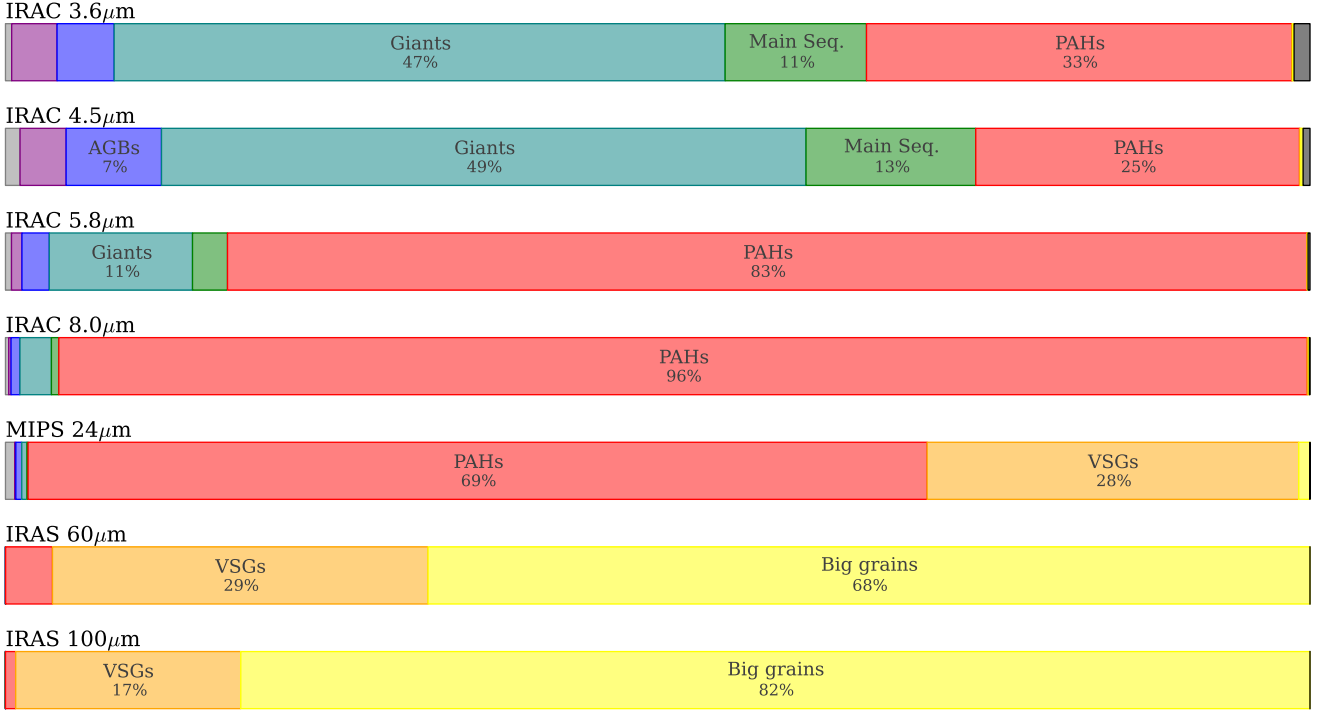}
\caption{The fractional contribution of various stellar populations and dust grain sizes to the total flux in the survey area as a function of wavelength. The colors are as in Figure~\ref{fig:spectra}. Only the main contributing spectral types and dust types are labeled.\label{fig:flux_contributions}}
\end{center}
\end{figure*}

\begin{figure*}
\begin{center}
\includegraphics[width=\textwidth]{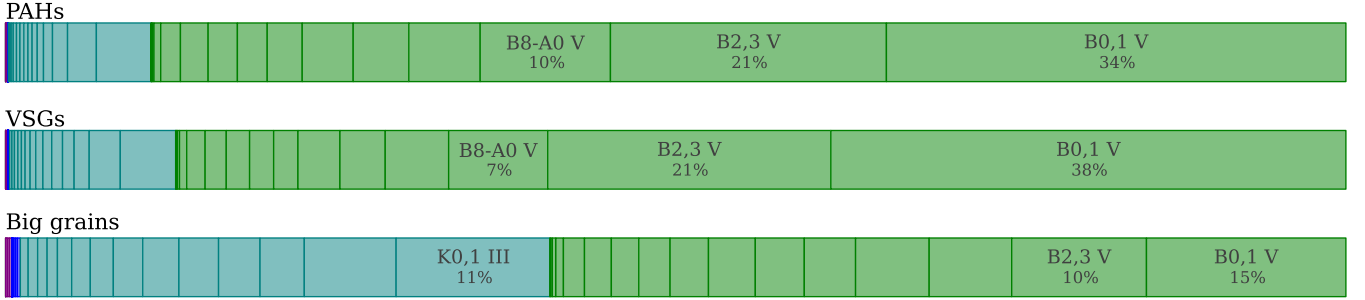}
\caption{The fractional contribution to dust heating as a function of individual spectral type. The colors are as in Figure~\ref{fig:spectra}, and the darker vertical lines indicate separations between individual spectral types. Only the main contributing spectral types are labeled.\label{fig:dust_contributions}}
\end{center}
\end{figure*}

\section{Analysis}

\label{sec:analysis}

In this section, we use the improved model as described in \S\ref{sec:improved}, but we note that the qualitative results in \S\ref{sec:contributions}
and \S\ref{sec:unresolved} are not significantly changed if we use the initial model instead.

\subsection{Flux and energy breakdown}

\begin{figure*}
\begin{center}
\includegraphics[width=0.75\textwidth]{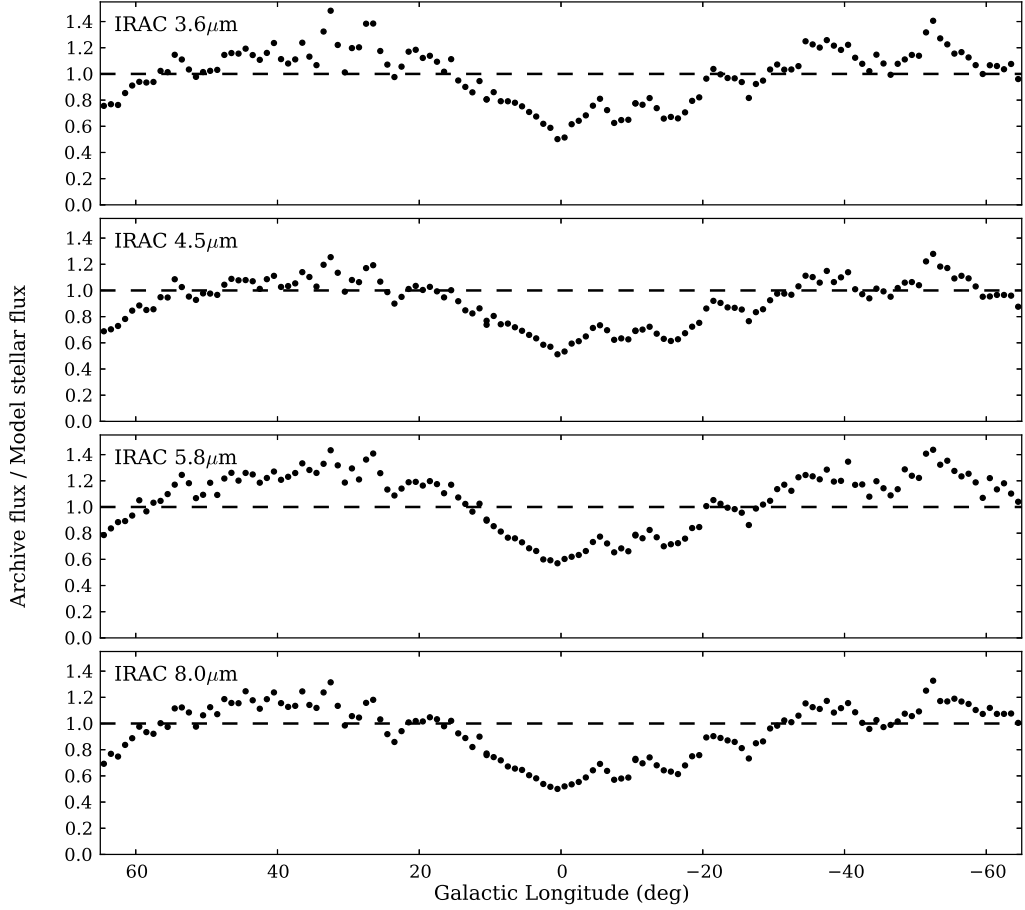}
\caption{The ratio of stellar emission measured in the GLIMPSE point source Archives to the total stellar flux predicted by the final model.  This suggests that unresolved stellar flux contributes a large component of the total stellar flux inside $|\ell|<20^\circ$. \label{fig:unresolved}}
\label{default}
\end{center}
\end{figure*}

\label{sec:contributions}

Figures~\ref{fig:spectra} to ~\ref{fig:morepah_profiles} show the breakdown of the surface brightness into contributions from the different components of the models, split up into main-sequence stars, giants, AGB stars, supergiants, other spectral classes, the three dust size ranges, and scattered light. We summarize this information into a more intuitive representation in Figure~\ref{fig:flux_contributions}. Approximately half of the IRAC 3.6 and 4.5\microns flux is from giants, with the second and third biggest contributors being PAHs and main-sequence stars respectively. On the other hand, the IRAC 5.8 and 8.0\microns bands are strongly dominated by emission from PAHs. MIPS 24\microns is dominated by emission two thirds from PAHs (in the form of continuum emission rather than features) and one third from VSGs, while IRAS 60 and 100\microns are dominated by thermal emission from the larger dust grains.

A similar analysis can be done to understand which stellar populations contribute the most to the energy injected into and reprocessed by the dust populations. Figure~\ref{fig:dust_contributions} shows the breakdown of this energy by individual spectral type (with the main contributing spectral types indicated). Heating of the dust grains is dominated by B-type stars and to a lesser extent, OB associations. Since O stars are rarer and have shorter lifetimes, the heating from the O stars is likely to be less uniform, since O stars are mostly found in clusters. In fact, the strong peaks in emission seen in the longitude profiles, which correspond to regions of massive star formation, are likely to be heated by a larger fraction of O-type stars. For the larger dust grains, which are less sensitive to UV radiative than the PAHs and VSGs, giants provide around a third of the heating.

\subsection{Unresolved flux}

\label{sec:unresolved}

The model derived in \S\ref{sec:improved} can be used to understand how much of the diffuse emission in the IRAC bands is due to unresolved stellar flux. In order to study this, we first computed the mean surface brightness of point sources in the GLIMPSE Archive source lists in $1^\circ$ bins in longitude, and compared this to the model stellar flux (excluding the dust emission). Figure \ref{fig:unresolved} shows the ratio of stellar flux measured in the GLIMPSE Archives compared to the model flux. For $|\ell| > 20^\circ$, the agreement between the resolved stellar flux and the model is good, suggesting that there is little or no unresolved stellar flux at these longitudes. Inside $|\ell| < 20^\circ$, the observed fluxes are as low as half of the predicted flux in places, suggesting that in that longitude range, the unresolved stellar flux could be as important a component to the total stellar flux as the resolved stellar flux.

\subsection{External viewpoint}

\label{sec:external}

\begin{figure*}
\begin{center}
\includegraphics[width=0.45\textwidth]{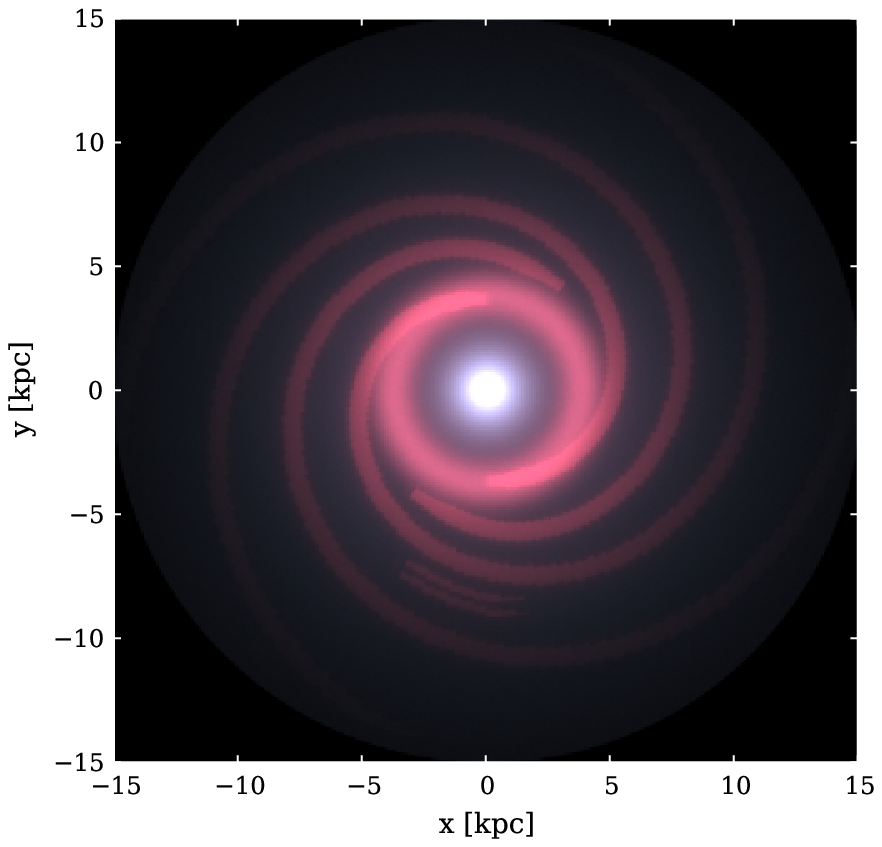}
\includegraphics[width=0.45\textwidth]{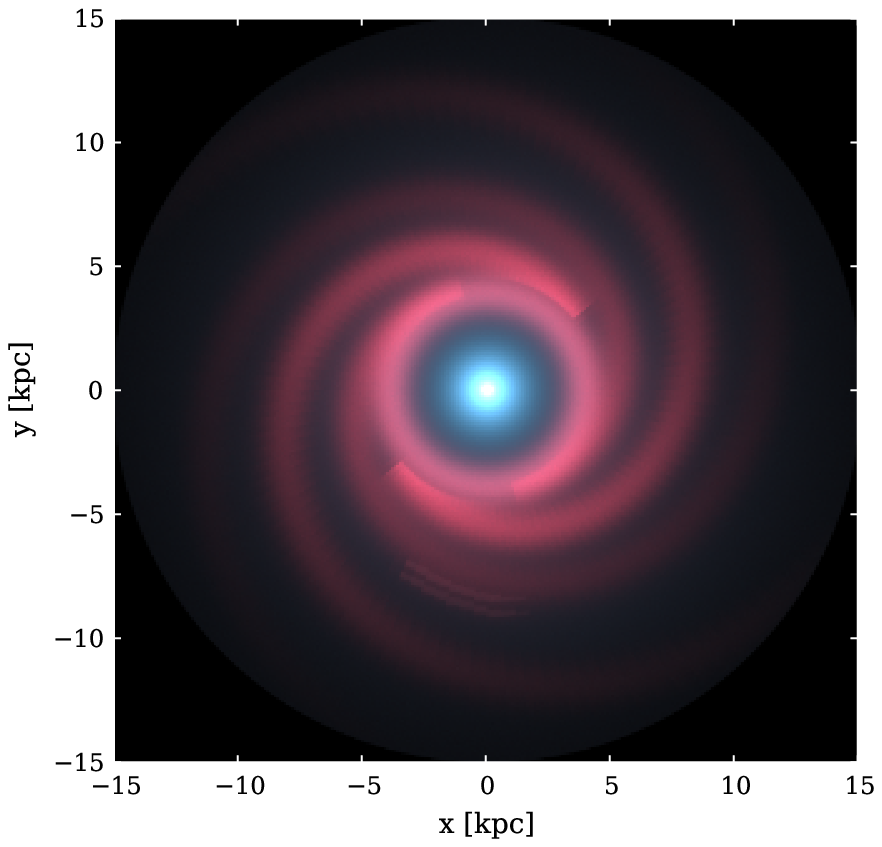}
\caption{A color composite view of the Galaxy model from an external viewpoint, viewed along the North Galactic Pole, for the initial model (left) and the final modified model (right). The Sun is located at (x, y) = (0, $-$8.5\kpc). The colors show IRAC 8.0\microns (red), IRAC 4.5\microns (green), and IRAC 3.6\microns (blue), all on a square-root intensity scale. \label{fig:external}}
\label{default}
\end{center}
\end{figure*}

As well as computing the appearance of the model from the position of the Sun, we can also compute the appearance of the Galaxy from an external viewpoint. In Figure \ref{fig:external}, we show what the model would look like at IRAC wavelengths viewed along the North Galactic pole. The spiral arms are prominent in PAH emission, while the bulge is bright in IRAC 3.6\microns and 4.5\microns. The molecular ring also features prominently, although as noted previously, there is debate as to whether this is in fact a real component of our Galaxy, or whether it is simply formed by an inward extension of the major spiral arms \citep[see e.g.][]{dobbs:12:2940}.

\section{Summary}

\label{sec:summary}

A radiative transfer model of the Galaxy which uses the SKY model in conjunction with  the dust properties from \citet{draine:07:810} was developed to self-consistently calculate the heating of dust grains, and verify whether it is able to reproduce the observed surface brightness from 3.6\microns to 100\microns. The main findings presented in this paper are the following:

\begin{enumerate}
\item The initial model is able to roughly reproduce the order of magnitude of the observed surface brightnesses observed, but there are disagreements between the model and observations, notably for $|\ell| > 30^\circ$ for the IRAC and MIPS bands, and inside $|\ell| < 30^\circ$ for IRAS 60\microns and 100\microns.
\item By modifying the model to incorporate two major stellar spiral arms and two secondary spiral arms with only young massive stars, as well as removing dust from the central few kpc of the Galaxy, we are able to significantly improve the quality of the fit. A slight systematic offset remains at IRAC 5.8 and 8.0\microns. By increasing the abundance of PAHs by 50\%, we are able to eliminate this systematic offset, though we caution that other effects may explain this offset.
\item The flatness of the IRAC 5.8 and 8.0\microns and MIPS 24\microns emission is not directly due to a hole in the dust distribution, but is a consequence of the lack of strong UV sources within the inner few kpc of the Galactic center. The evidence for a deficit of dust is only apparent at longer wavelengths, in the IRAS 60 and 100\microns bands.
\item Since the stellar populations are not symmetrically distributed, the heating of the dust is not uniform, and therefore the spiral arms appear in diffuse emission at infrared wavelengths without the need for a non-symmetrical dust distribution (i.e. the model does not require additional dust in the spiral arms). This is not to say that the distribution of dust is not enhanced in the spiral arms, but that there is no evidence from the data that it is.
\item The overall flux at IRAC 3.6 and 4.5\microns is dominated by giant stars, PAHs, and main-sequence stars, while the flux from IRAC 5.8\microns to IRAS 100\microns probes mostly dust, with the size of the dust grains probed increasing with wavelength.
\item On large scales, transiently heated very small grains and PAH molecules are predominantly heated by B-type stars. The larger dust grains also have a significant component of heating from giant stars.
\item The \textit{Spitzer}/IRAC bands may contain as much unresolved stellar flux as resolved stellar flux for $|\ell| < 20^\circ$.
\end{enumerate}

The models presented here, while simple, help us quantify the main factors that determine the large-scale Galactic mid-plane emission. We plan to carry out significant improvement to the models in the future, including a more systematic exploration of parameter space, a more realistic clumpy distribution of dust, and an improved treatment of the inner Galaxy, specifically relating to the molecular ring and the Galactic bar. In addition, the modeling of observations can be extended to spatial and wavelength regions outside that covered here. For instance, one could include observations for the outer Galaxy (GLIMPSE 360) or at longer wavelengths (\textit{Herschel} HiGal). Once the data from the Wide-Field Infrared Survey Explorer (WISE) are fully released, it will even be possible to model the all-sky observations from 3.5\microns to 100\microns when combined with the IRAS all-sky data.

\section*{Acknowledgments}

We wish to thank the referee for a careful review and for comments that helped improve this paper. We thank Bruce Draine and Aigen Li for useful discussions relating to PAH emission. This work is based in part on observations made with the Spitzer Space Telescope, which is operated by the Jet Propulsion Laboratory, California Institute of Technology under a contract with NASA. TR was supported by NASA through the Spitzer Space Telescope Fellowship Program, and BW, EC, MM, and BB gratefuly acknowledge  support from NASA through awards issued by JPL/Caltech (1368699 and 1367334). The models in this paper were run using Hyperion 0.9.0, and the scripts used to produce the results presented here are available as supplementary online material.

\bibliography{}

\end{document}